# Shirokov's contracting lifetimes and the interpretation of velocity eigenstates for unstable quantons[1]

Gordon N. Fleming*
*Prof. Emeritus, Physics, Pennsylvania State Univ.*
* gnf1@psu.edu , gnf1@earthlink.net

This paper is concerned with the interpretation of velocity eigenstates for unstable quantons[1], their relationship to space-like momentum eigenstates for such quantons and the explanation of Shirokov's contracting lifetimes for such velocity eigenstates. It is an elaboration of a portion of the authors earlier study [F 09].

**1. Introduction:** In 2006 Shirokov [S 06] presented a derivation that the rate of decay for an unstable quanton with a precise, non-zero, velocity was required by Lorentz covariance to be *faster* than the rate for zero velocity. Taken at face value this result seems to contradict the traditional view that time dilation for moving quantons would slow down the decay rate. That traditional view is motivated in part by the expectation that the time dilation with increasing velocity of classical unstable *particles* would carry over into quantum theory. Shirokov did point out that *momentum* eigenstates, as opposed to *velocity* eigenstates, do display the expected slow down. But whence the discrepancy and what is its significance? Hegerfeldt [H 06] simplified Shirokov's derivation and indicated that the result was sufficiently counterintuitive to demand further study. More recently Shirokov provided further discussion of his result [S 08].

In a recent survey of conflicting views in the theory of unstable quanton decay, [F 09] I argued that the apparent problematic character of Shirokov's result arose from a *misinterpretation* of velocity eigenstates of unstable quantons. For the puzzle arises initially only as a consequence of interpreting the velocity eigenstate as a possible state for *a quanton which is definitely undecayed at some instant of time*. If we then ask what time displacement of the state will reduce the inner product with the original state by a given factor, Shirokov shows us that the required time displacement decreases as the velocity increases, and by just the reciprocal of the factor which gives the conventional time dilation. But, in fact, the proposed interpretation of the state is admissible only for zero velocity! For non-zero velocity, as will be shown below, the quanton can be definitely undecayed *only on a non-instantaneous, space-like hyperplane*, the orientation of which in Minkowski space-time is determined by the velocity eigenvalue. We will call this hyperplane the *no-decay hyperplane* for the quanton. In [S 08] Shirokov recognized that non-zero velocity eigenstates can not be 100% undecayed at any instant of time, but he did not identify where the 100% condition *would* be satisfied. We will see here that, as shown in [F 09], if we consider two parallel hyperplanes, one of which is the no-decay hyperplane and the other is separated from the first by a time-like interval, orthogonal to the hyperplanes and of magnitude equal to the rest frame lifetime, then the *time interval* between these hyperplanes is just the contracted lifetime given by Shirokov's derivation (**Fig. 1**).

---

1. I follow Levy-Leblond [L 88] and others in referring to the molecular, atomic, nuclear and sub-nuclear constituents of the mass-energy world as quantons rather than particles.

That this equality is not just a coincidence can be seen as follows (details in **3**). All velocity eigenstates are states for which some set of three, mutually orthogonal, space-like components of the total 4-momentum are exactly zero. For an unstable quanton the remaining, orthogonal, time-like, component of the total 4-momentum carries the non-trivial mass distribution of the quanton and generates the dynamical decay. The no-decay hyperplane of the quanton must contain the space-like

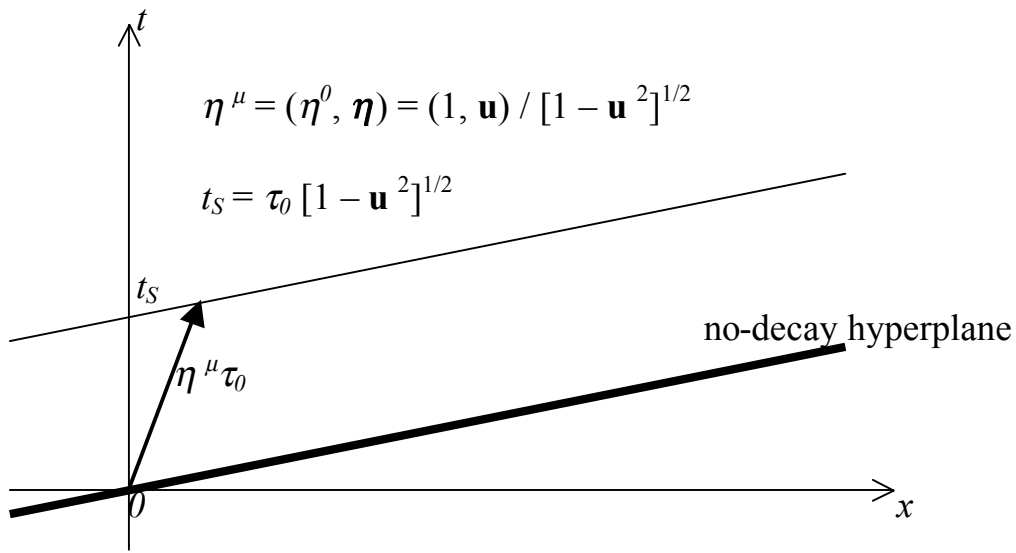

$$\eta^\mu = (\eta^0, \boldsymbol{\eta}) = (1, \mathbf{u}) / [1 - \mathbf{u}^2]^{1/2}$$

$$t_S = \tau_0 [1 - \mathbf{u}^2]^{1/2}$$

**Fig. 1:** Shirokov's lifetime, $t_S$, for an unstable quanton with velocity, **u**, is the *time* interval separating the no-decay hyperplane from a later, parallel hyperplane separated by the orthogonal *time-like* interval, $\eta^\mu \tau_0$, with the magnitude of the rest frame lifetime, $\tau_0$.

directions corresponding to the zero momentum eigenvalues, for this is the only hyperplane orientation within which all translations fail to induce decay. For zero velocity these space-like directions are just the *spatial* directions and the no-decay hyperplane is instantaneous. For non-zero velocity the zero momentum, space-like directions are not *all* spatial directions and, therefore, the no-decay hyperplane is non-instantaneous. Denoting the ratio of the velocity eigenvalue to the vacuum speed of light by **u** and the dimensionless, time-like, unit 4-vector orthogonal to the no-decay hyperplane by $\eta$, we have (using the $+---$ metric),

$$\mathbf{u} = \vec{\eta}/\eta^0, \tag{1.1a}$$

$$\eta^\mu = (\eta^0, \vec{\eta}) = (1, \mathbf{u})/\sqrt{1-\mathbf{u}^2}. \tag{1.1b}$$

The attractive feature of velocity eigenstates for modeling unstable quantons with unsharp mass spectra, is that, unlike 3-momentum eigenstates, they transform into each other under the homogeneous Lorentz group. It was on this basis in [S 08] that Shirokov recognized the presence of decay products at *all* instants of time in the non-zero velocity eignstates. For the Lorentz boost generators include contributions from the decay producing interactions [F 02]. Accordingly, the no-decay hyperplane is always instantaneous only in precisely that inertial frame in which the transformed velocity is zero. With this understanding the velocity eigenstates can not constitute a complete basis for unstable quantons since, (1) for any given no-decay hyperplane, only one velocity eigenvalue can occur among the basis states for that hyperplane and (2) any superposition of velocity eigenstates with distinct eigenvalues will not consist of a pure, undecayed parent on *any* hyperplane.

A complete basis for an unstable quanton with a given no-decay hyperplane *is* contained in that class of states which have non-zero, as well as zero, eigenvalues for the space-like components of the total 4-momentum that are parallel to the hyperplane . We will call such states space-like momentum (SLM) eigenstates. It remains the case that the no-decay hyperplane must contain the directions of the diagonalized SLM components since translations along those directions induce, at most, phase factors, but no decay. As before, decay is generated by the time-like component of the total 4-momentum orthogonal to the no-decay hyperplane, which carries the non-trivial energy-like distribution due now to the mass distribution augmented by the non-zero SLM eigenvalues. But now these SLM eigenstates can span the state space for a given no-decay hyperplane. We will discuss the result, alternatively derived in [S 06 ] and [F 09], that these SLM eigenstates do display lifetime dilation, albeit modified by the spread in the mass spectrum.

**2. Velocity eigenstates:** We take our system to be isolated, i.e., to have conserved total 4-momentum operators, $\hat{P}^\mu$. The velocity operator is given by, $\hat{\mathbf{P}}/\hat{P}^0$. A velocity eigenstate then, $|\mathbf{u},\alpha\rangle$, (where $\alpha$ represents any other properties, such as spin, that may be carried by the system) is defined by,

$$(\hat{\mathbf{P}}/\hat{P}^0)|\mathbf{u},\alpha\rangle = |\mathbf{u},\alpha\rangle\mathbf{u}, \qquad (2.1a)$$

or,

$$\hat{\mathbf{P}}|\mathbf{u},\alpha\rangle = \hat{P}^0|\mathbf{u},\alpha\rangle\mathbf{u}. \qquad (2.1b)$$

With a little manipulation, using (1.1), we obtain from (2.1b),

$$(\hat{P}^\mu - \eta^\mu(\eta\hat{P}))|\mathbf{u},\alpha\rangle = 0, \qquad (2.2)$$

and our velocity eigenstate has zero space-like momentum in all directions orthogonal to $\eta^\mu$.

Since we can also recover (2.1) from (2.2), using (1.1), the zero space-like momentum condition is equivalent to the velocity eigenstate condition, i.e., a state is a velocity eigenstate iff it is a zero momentum eigenstate for all space-like directions orthogonal to a time-like direction determined by the velocity eigenvalue.

Defining the mass operator, $\hat{M}$, by (restricting to non-negative mass spectra),

$$\hat{M} := \left|\sqrt{\hat{P}^2}\right|, \qquad (2.3)$$

and assuming (2.2), we have,

$$\eta\hat{P}|\mathbf{u},\alpha\rangle = \hat{M}|\mathbf{u},\alpha\rangle. \qquad (2.4)$$

Consequently, if $|\mathbf{u},\alpha\rangle$ is a *mass* eigenstate, then it is a 4-momentum eigenstate as well, with 4-momentum in the direction of $\eta^\mu$. On the other hand, if $|\mathbf{u},\alpha\rangle$ represents an unstable quanton, then the decay process requires the mass spectrum to be spread and only the space-like components of $\hat{P}^\mu$ orthogonal to $\eta^\mu$ are precise, being zero.

**3. Unstable quantons:** A physical system can consist of *only* one undecayed, unstable quanton on, at most, one space-like hyperplane. To be so constituted on two parallel hyperplanes, in the presence of time-like reversal invariance, would require a periodic evolution rather than decay. Similarly, to be so constituted on two intersecting hyperplanes, in the presence of Lorentz invariance, would require an alternative form of periodic evolution rather than decay. But what we mean here by an *unstable quanton* is a quanton that spontaneously evolves into a superposition of parent and decay products and which does not, subsequently or elsewhere, evolve back into a pure parent. The one preferred hyperplane on which no decay products are to be found for the isolated unstable quanton we call the *no-decay hyperplane*. We now argue, in more detail than above, that for such a quanton in a velocity eigenstate with non-zero eigenvalue, the no-decay hyperplane can not be instantaneous.

From (2.2) any active translation orthogonal to $\eta^\mu$ changes nothing, i.e., for any translation vector, $a^\mu$, satisfying, $a^\mu = a^\mu - \eta^\mu(\eta a)$, we have,

$$\exp[(i/\hbar)a\hat{P}]|\mathbf{u},\alpha> = |\mathbf{u},\alpha>. \tag{3.1}$$

Consequently the unstable quanton can decay only *due* to evolution along the time-like direction, $\eta^\mu$.

Now all active translations *within*, i.e., parallel to the no-decay hyperplane can only slide the quanton around within that hyperplane. They can not induce decay. But a purely *spatial* translation with a non-zero component in the direction of **u**, i.e., the direction of $\bar{\eta}$, *will* induce decay since, from (1.1) and (2.2),

$$\hat{\mathbf{P}}|\mathbf{u},\alpha> = \mathbf{u}\eta^0(\eta\hat{P})|\mathbf{u},\alpha>, \tag{3.2a}$$

and such a translation would be *identical* to one in the time-like direction of $\eta^\mu$. This kind of argument applies to any hyperplane that is not orthogonal to $\eta^\mu$, and thus the no-decay hyperplane for the state, $|\mathbf{u},\alpha>$, must be orthogonal to $\eta^\mu$.

From this, then, it follows that, for **u** ≠ **0**, the no-decay hyperplane for the quanton can not be instantaneous.

Furthermore, from,

$$\hat{P}^0 | \mathbf{u},\alpha \rangle = \eta^0 (\eta\hat{P}) | \mathbf{u},\alpha \rangle, \tag{3.2b}$$

which also follows from (1.1) and (2.2), we have that a pure *time* evolution through the interval, $t_S$, is *identical* to a time-like evolution, in the direction of $\eta^\mu$, of magnitude, $\tau_0$, the rest frame lifetime. Thus,

$$\begin{aligned} \exp[(i/\hbar)t_S \hat{P}^0] | \mathbf{u},\alpha \rangle &= \exp[(i/\hbar)t_S \eta^0 (\eta\hat{P})] | \mathbf{u},\alpha \rangle \\ &= \exp[(i/\hbar)\tau_0 (\eta\hat{P})] | \mathbf{u},\alpha \rangle \end{aligned}, \tag{3.3}$$

and we have Shirokov's result,

$$t_S = \tau_0 / \eta^0 = \tau_0 \sqrt{1 - \mathbf{u}^2} \leq \tau_0, \tag{3.4}$$

consistently interpreted.

Thus the velocity eigenstates for an unstable quanton can not be a basis for such a system since no two with distinct eigenvalues can even share the same no-decay hyperplane. To span the space of unstable quanton states having a *given* no-decay hyperplane, we can, instead, employ the SLM eigenstates for all space-like momentum components parallel to the no-decay hyperplane.

**4. Space-like momentum eigenstates:** We call a state, $|p,\eta,\alpha\rangle$, a space-like momentum (SLM) eigenstate, iff there exists a future pointing time-like unit vector, $\eta^\mu$, such that,

$$(\hat{P}^\mu - \eta^\mu(\eta\hat{P})) | p,\eta,\alpha \rangle = | p,\eta,\alpha \rangle p^\mu \tag{4.1}$$

Clearly, $\eta p = 0$, and the relation to the velocity eigenstates is given by,

$$|\mathbf{u},\alpha\rangle \propto |0,\eta,\alpha\rangle, \tag{4.2}$$

with the proportionality determined by the normalization conventions adopted for these states and with (1.1) relating **u** and $\eta$. By fixing $\eta$ and varying $p$ (and, possibly, $\alpha$) we can stay with one hyperplane and span the states of the unstable quanton on that no-decay hyperplane. Henceforth we will make the no-decay hyperplane explicit by expanding the notation, $\eta$, to that of ($\eta$, $\tau$) denoting the no-decay hyperplane as that containing the points with Minkowski coordinates, $x$, satisfying $\eta x = \tau$.

For any unit norm state of such a quanton, $|\Psi;\eta,\tau>$, we can extract the contributing SLM states by translation and superposition, i.e.,

$$(2\pi\hbar)^{-3}\int d^4\lambda\, \delta(\eta\lambda)\exp[(i/\hbar)\lambda(\hat{P}-p)]|\Psi;\eta,\tau>$$

$$=\delta_\eta^3(\hat{P}-p)|\Psi;\eta,\tau>=\sum_\alpha \int |p;\eta,\tau;\alpha><p;\eta,\tau;\alpha|\Psi;\eta,\tau>, \qquad (4.3)$$

where,

$$<p';\eta,\tau;\alpha'|p;\eta,\tau;\alpha>=\delta_\eta^3(p'-p)\delta(\alpha',\alpha). \qquad (4.4)$$

The projection onto the state space of the single, unstable quanton with no-decay hyperplane, ($\eta,\tau$), is given by,

$$\hat{\Pi}(\eta,\tau):=\sum_\alpha \int\int d^4p\,\delta(\eta p)|p;\eta,\tau;\alpha><p;\eta,\tau;\alpha|. \qquad (4.5)$$

For fixed $\eta,\tau;\alpha$ the individual SLM eigenstates, $|p;\eta,\tau;\alpha>$, are connected by the unitary transformations generated by the generalized Newton-Wigner position operator [F 65, 99] on the ($\eta,\tau$) hyperplane, $\hat{Q}^\mu(\eta,\tau)$, where, $\eta\hat{Q}(\eta,\tau)=0$. Thus it follows from the commutation relations,

$$[\hat{Q}^\mu(\eta,\tau),\hat{Q}^\nu(\eta,\tau)]=0, \qquad (4.6a)$$

$$[\hat{Q}^\mu(\eta,\tau),\hat{K}_\nu(\eta)]=-i\hbar(\delta_\nu^\mu-\eta^\mu\eta_\nu), \qquad (4.6b)$$

and,

$$[\hat{Q}^\mu(\eta,\tau),\eta\hat{P}]=i\hbar\frac{\hat{K}^\mu(\eta)}{\eta\hat{P}}, \qquad (4.6c)$$

where,

$$\hat{K}^\mu(\eta) := \hat{P}^\mu - \eta^\mu(\eta\hat{P}) \quad \text{and} \quad \eta\hat{P} = \sqrt{\hat{M}^2 - \hat{K}(\eta)^2}, \tag{4.6d}$$

that,

$$\exp[(i/\hbar)(p-p')\hat{Q}(\eta,\tau)]|p;\eta,\tau;\alpha\rangle = |p';\eta,\tau;\alpha\rangle \tag{4.7}$$

($\hat{M}$ having been defined in (2.3)). We furthermore have,

$$\exp[(i/\hbar)\sqrt{\hat{M}^2 - \hat{K}(\eta)^2}\,\tau]\exp[-(i/\hbar)\hat{Q}(\eta,0)p]$$
$$= \exp[-(i/\hbar)\hat{Q}(\eta,0)p]\exp[(i/\hbar)\sqrt{\hat{M}^2 - (\hat{K}(\eta)+p)^2}\,\tau], \tag{4.8}$$

which, in turn yields,

$$|p;\eta,\tau;\alpha\rangle = \exp[(i/\hbar)\eta\hat{P}\tau]|p;\eta,0;\alpha\rangle$$
$$= \exp[-(i/\hbar)\hat{Q}(\eta,0)p]\exp[(i/\hbar)\sqrt{\hat{M}^2 - p^2}\,\tau]|0;\eta,0;\alpha\rangle. \tag{4.9}$$

This last enables us to compare the rate of decay for different values of $p$. As detailed in [F 09 (see section **4** and Appendix **1**)], defining the lifetime, $T_{p,\alpha}$, by,

$$T_{p,\alpha} := \int_0^\infty d\tau\, |I_{p,\alpha}(\tau)|^2, \tag{4.10a}$$

where,

$$\langle p';\eta,\tau;\alpha'|p;\eta,0;\alpha\rangle := \delta_\eta^3(p'-p)\delta(\alpha',\alpha)I_{p,\alpha}(\tau), \tag{4.10b}$$

then from,

$$I_{p,\alpha}(\tau) = \int d\mu\, \sigma_\alpha(\mu)\exp[-(i/\hbar)\sqrt{\mu^2 - p^2}\,\tau], \tag{4.10c}$$

we find,

$$T_{p,\alpha} = \pi\hbar \int d\mu\, \sigma_\alpha(\mu)^2 \frac{\sqrt{\mu^2 - p^2}}{\mu}, \tag{4.10d}$$

(dependence on $\alpha$ can occur if $\alpha$ varies over different *types* of unstable quantons).

With (4.10d) we clearly see the relativistic *dilation* of lifetime with increasing $-p^2$, albeit modulated from the sharp classical result by the indefinite, i.e., spread, mass spectrum of the parent quanton.

**5. Discussion:** Classically, the time dilation of unstable *particle* lifetimes seems more closely tied to the kinematical concept of velocity than the dynamical concept of momentum. How is it that the transition from classical unstable particles to unstable quantons shifts the burden of lifetime dilation to momentum? The answer is ultimately traceable to the energy-time uncertainty relation which forces the spread in the mass spectrum of the unstable quanton and thereby renders the velocity eigenstates to be only a small subclass of the SLM eigenstates required to form bases of states for given no-decay hyperplanes.

There is ,of course, nothing wrong with inquiring into the *time* interval lying between the no-decay hyperplane and the parallel unit-lifetime hyperplane for velocity eigenstates. But regarding the contracted result as counterintuitive results from neglecting the non-instantaneity of those hyperplanes. Relative to their own no-decay hyperplane, all velocity eigenstates are at rest with rest frame lifetimes and, consequently, no superposition of velocity eigenstates with distinct velocity eigenvalues can yield unit probability for finding only the parent quanton anywhere or anywhen in any inertial frame. Only the SLM eigenstates with non-zero momenta have non-zero velocity relative to the frames in which their own no-decay hyperplanes are instantaneous. But *those* velocities are all indefinite, not sharp, i.e.,

$$\frac{\hat{K}^\mu(\eta)}{\eta\hat{P}}\mid p;\eta,\tau;\alpha> = \mid p;\eta,\tau;\alpha> p^\mu <(\eta\hat{P})^{-1}>_{p,\alpha} + \mid p;\eta,\tau;\alpha_{K(\eta)/\eta P}> \Delta_{p,\alpha}\left(\frac{\hat{K}^\mu(\eta)}{\eta\hat{P}}\right), \quad (5.1)$$

where, $\mid p;\eta,\tau;\alpha_{K(\eta)/\eta P}>$, contains nothing but decay products on the $(\eta,\tau)$ hyperplane. The lifetime, (4.10d), does increase with the squared magnitude of the velocity expectation value, $-p^2 < (\eta\hat{P})^{-1} >^2_{p,\alpha}$, but not so familiarly as the dependence on $-p^2$ indicated by (4.10d). If we furthermore look at the instantaneous 3-velocity of the SLM eigenstates we find ,

$$<\frac{\hat{\mathbf{P}}}{\hat{P}^0}>_{p,\alpha} = <\frac{\mathbf{p}+\vec{\eta}(\eta\hat{P})}{p^0+\eta^0(\eta\hat{P})}>_{p,\alpha} = <\frac{\mathbf{p}+\vec{\eta}(\eta\hat{P})}{(\vec{\eta}\mathbf{p}/\eta^0)+\eta^0(\eta\hat{P})}>_{p,\alpha}, \quad (5.2)$$

even less closely related, formally, to *classical* lifetime dilation.

Finally, let us note in passing that if one examines the velocity eigenstate matrix elements of the projection operator defined in (4.5) and evaluated on instantaneous hyperplanes, i.e., at definite times, $<\mathbf{u}',\alpha'|\hat{\Pi}(t)|\mathbf{u},\alpha>$, one finds non-trivial time dependence only for the case $\mathbf{u} = \mathbf{u}' = \mathbf{0}$ and when $\mathbf{u}$ and $\mathbf{u}'$ are collinear, but unequal. However, for $\mathbf{u}' = \mathbf{u} =/= \mathbf{0}$, and for non-collinear velocities, the matrix element is time *independent*! This is essentially unintelligible if one analyses the matter solely from the perspective of structure and content at definite times. But it is easily understood, along lines laid out in the **Appendix**, below, in terms of the *intersecting* no-decay hyperplanes of the velocity eigenstates and the instantaneous projection operator.

For classical, unstable *particles* there is no question of a no-decay hyperplane or a distinction between sharp velocity and momentum. They live on world lines, just as stable particles do, albeit spontaneously terminating ones. The indefinite mass of unstable *quantons* and the unitary nature of the decay evolution (prior to *detection* of decay) sever the connection between sharp velocity and sharp momentum and allow, at most, one hyperplane devoid of decay products. Furthermore, 'though we have not touched on it here, the operator character and hyperplane dependence of the global positions for quantons, stable or unstable, preclude any sharp notion of a worldline. The quantum world is far richer than the classical and many comfortable features of the latter simply do not carry over. As Shirokov showed, lifetime dilation with sharp velocity for unstable quantons is one such.

**Appendix: Time-like dependencies of mixed matrix elements of the unstable quanton state space projectors**

Having established the interpretation of the SLM eigenstates, $|p;\eta,\tau;\alpha>$, it is illuminating, and perhaps a little surprising, to examine the dependencies on time-like parameters of the mixed matrix elements of the state space projectors (defined in (4.5)), $\hat{\Pi}(\eta,\tau)$, for unstable quantons on given no-decay hyperplanes. By *mixed* matrix elements we mean matrix elements in which the SLM eigenstate ket and bras need not belong to the same basis, i.e., they may correspond to distinct no-decay hyperplanes. The general example is of the form,

$$<p';\eta',\tau';\alpha'|\hat{\Pi}(\eta'',\tau'')|p;\eta,\tau;\alpha>, \qquad (A.1)$$

and we will examine the dependencies on $\tau$, $\tau'$ and $\tau''$, which dependencies will be found to depend on the relationships between $\eta$, $\eta'$ and $\eta''$.

While we have not provided formally exhaustive and precise definitions of the SLM eigenstates we are discussing, we here provide enough information concerning them to determine the dependencies of interest. In particular we require,

$$\exp[(i/\hbar)a\hat{P}]|p;\eta,\tau;\alpha> = |p;\eta,\tau+a\eta;\alpha>\exp[(i/\hbar)ap], \quad (A.2)$$

for arbitrary translation vector, $a^\mu$. Although we will not use it here, we also stipulate,

$$\exp[-(i/2\hbar)\hat{M}\omega]|p;\eta,\tau;\alpha> = |\Lambda(\omega)p;\Lambda(\omega)\eta,\tau;\alpha_\omega>, \quad (A.3)$$

where, $\hat{M}^{\mu\nu}$, is the hermitian generator of the homogeneous Lorentz transformations, $\Lambda(\omega)^\mu_\nu = (e^\omega)^\mu_\nu$, are the coefficients for the Lorentz transformation of tensors and the little $\omega$ subscript on $\alpha$ covers a multitude of sins.

If we now insert the identity operator in the form,

$$\hat{I} = \exp[-(i/\hbar)a\hat{P}]\exp[(i/\hbar)a\hat{P}], \quad (A.4)$$

into (A.1) between the left hand bra and the projector and between the projector and the right hand ket, then using (A.2), its' conjugate form and (4.5) we obtain,

$$<p';\eta',\tau';\alpha'|\hat{\Pi}(\eta'',\tau'')|p;\eta,\tau;\alpha> =$$

$$\exp\left[-\frac{i}{\hbar}ap'\right]<p';\eta',\tau'+a\eta';\alpha'|\hat{\Pi}(\eta'',\tau''+a\eta'')|p;\eta,\tau+a\eta;\alpha>\exp\left[\frac{i}{\hbar}ap\right], \quad (A.5)$$

a relationship fraught with implications!

Case 1: Let us begin our analysis of (A.5) by considering the case in which $\eta$, $\eta'$ and $\eta''$ are all distinct and linearly independent. In that case there is one space-like direction such that any $a$ parallel to it is orthogonal to all the $\eta$ vectors. Considering such $a$ in (A.5) requires $(p – p')a = 0$ if the matrix element is to be non-zero. If we then consider $a$ that are orthogonal to only two of the $\eta$ vectors (and any two can be so chosen) we find that dependence on the $\tau$ variable associated with the $\eta$ vector *not* orthogonal to $a$ is constrained to be at most in the form of a phase factor. So in this case all time-like dependencies are severely restricted to what will be called *trivial*, phase factor form.

Case 2: Next we consider the case in which exactly one linear relationship holds among the $\eta$ vectors. Let it be, $\lambda\eta + \lambda'\eta' + \lambda''\eta'' = 0$, with at least two of the coefficients different from zero. Then there is a *two* dimensional family of space-like directions orthogonal to all the $\eta$ vectors and for any $a$ parallel to that family we must have, $(p - p')a = 0$, for the matrix element to be non-zero. With this condition satisfied, no constraint is provided by (A.5) for the dependence on the variable, $\lambda\tau + \lambda'\tau' + \lambda''\tau''$, but all other combinations of the $\tau$ variables linearly independent of the preceding one can contribute to at most trivial dependence of the matrix element, as consideration of translations, $a$, outside the two dimensional family will show. Note that this case includes the situation where $\eta = \eta' =/= \eta''$, an unmixed matrix element.

Case 3: Finally we consider the case in which two, independent, linear relations hold among the $\eta$ vectors. Now there is a full three dimensional family of space-like directions orthogonal to all the $\eta$ vectors and the matrix element can be non-zero only when, $p - p' = 0$, and then (A.5) places no constraint for the dependence on the independent linear combinations of the $\tau$ variables corresponding to the linear relations among the $\eta$ vectors. In fact, the $\eta$ vectors all being future pointing time-like unit vectors, the existence of two, independent, linear relations among them requires they all be equal!

So (A.5) tells us a good deal about the time-like dependencies of the matrix element. But we might still be a little puzzled about the instances of restriction to merely phase factor dependence. A look at the geometrical relationships among the no-decay hyperplanes sheds some light on this issue (**Fig. A1, 2**).

In **Fig. A1** we have first the (top) situation in which two, independent, linear relationships hold among the $\eta$ vectors, resulting in equality of all the $\eta$ vectors and non-trivial dependence of the matrix element on the $\tau$ differences between the parallel no-decay hyperplanes, which differences measure changeable *global* relationships between pairs of hyperplanes. Second, we have the (bottom) situation in which only one linear relation holds among the $\eta$ vectors; a relation yielding equality between two of the $\eta$ vectors. Only the global relationship between the parallel hyperplanes is changeable and is measured by the $\tau$ difference between them on which the matrix element can depend non-trivially. The global

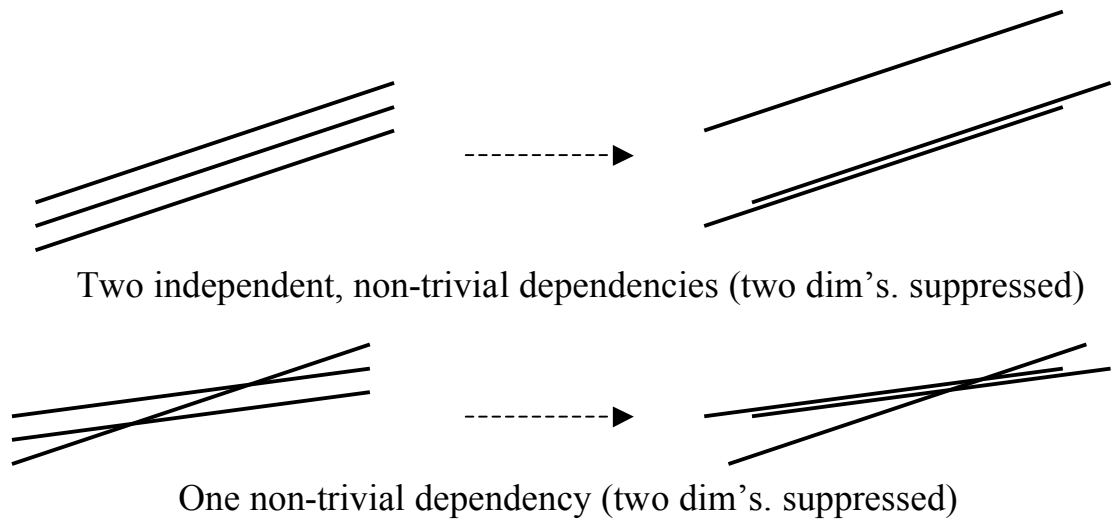

Two independent, non-trivial dependencies (two dim's. suppressed)

One non-trivial dependency (two dim's. suppressed)

**Fig. A1**

relationship between the third hyperplane and either of the parallel ones can not change; only the *location* of the intersections can be altered corresponding to the restriction to trivial dependence on the corresponding $\tau$ differences.

In **Fig. A2** we have first the (top) situation in which, again, one linear relation holds among the $\eta$ vectors, but this time no equality between any two of them. This allows the global relationship among the three hyperplanes to change with the size of the space-time region encompassed between them as measured by the corresponding linear combination of the $\tau$ variables on which the matrix element can depend non-trivially. The global relationship between any two of the hyperplanes can not change and, accordingly, dependence on the corresponding $\tau$ variable differences can only be trivial. Second, the (bottom) situation represents the absence of *any* linear relationship among the $\eta$ vectors and this results in an unavoidable common intersection of all three hyperplanes which can never be altered in form but only relocated. Thus the global relationships among the hyperplanes can not change with the $\tau$ differences and, accordingly, there are no non-trivial dependencies.

Ultimately, the source of the connection between global hyperplane relationships and trivial vs non-trivial dependence is the fact that SLM eigenstates are not located *anywhere* on their no-decay hyperplanes. They are simply on the hyperplanes *entire*, except, of course, for phase factors!

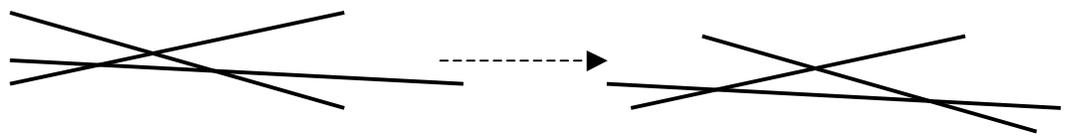

One non-trivial dependency (two dim's. suppressed)

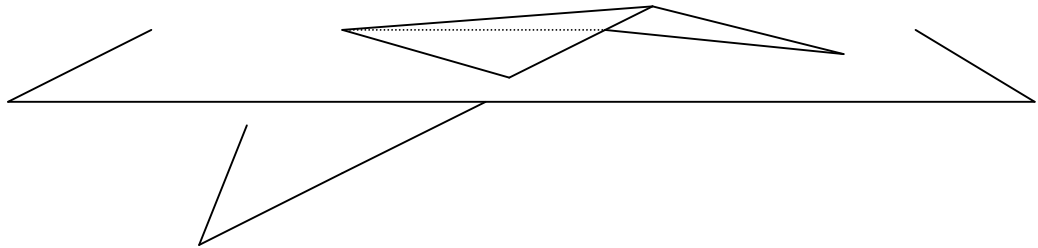

No non-trivial dependence possible (one dim. suppressed)

**Fig. A2**